\documentclass[prd,reprint,superscriptaddress,nofootinbib,longbibliography,aps]{revtex4-1}
\usepackage{graphicx}
\usepackage{longtable}
\usepackage{amsmath}
\usepackage{color}
\usepackage{dsfont}
\usepackage{amssymb}
\usepackage{subfigure}
\usepackage{tabularx}
\usepackage[colorlinks=true]{hyperref}
\usepackage{cleveref}
\usepackage[normalem]{ulem}
\usepackage{slashed}
\usepackage[toc]{appendix}
\usepackage{bm}
\usepackage{braket}
\usepackage{comment}

\begin{document}

\title{Apparent $w<-1$ and a Lower $S_8$ from Dark Axion and Dark Baryons Interactions}

\author{Justin Khoury}
\email{jkhoury@upenn.edu}
\affiliation{Center for Particle Cosmology, Department of Physics and Astronomy, University of Pennsylvania, Philadelphia, Pennsylvania 19104, USA}

\author{Meng-Xiang Lin}
\email{mxlin@upenn.edu}
\affiliation{Center for Particle Cosmology, Department of Physics and Astronomy, University of Pennsylvania, Philadelphia, Pennsylvania 19104, USA}

\author{Mark Trodden} 
\email{trodden@upenn.edu}
\affiliation{Center for Particle Cosmology, Department of Physics and Astronomy, University of Pennsylvania, Philadelphia, Pennsylvania 19104, USA}


\begin{abstract}
    We show that a simple coupling between dark energy and dark matter can simultaneously address two distinct hints at new physics coming from cosmological observations. The first is the recent evidence from the DESI project and supernovae observations that the dark energy equation of state~$w$ is evolving over cosmic time from an earlier value that is~$<-1$ to a present-day value~$>-1$. The second observation is the so-called~$S_8$ tension, describing the suppression of the growth of matter overdensities compared to that expected in the~$\Lambda$CDM model. We propose a stable, technically natural particle physics implementation of this idea, in which dark matter consists of dark baryons in a strongly-coupled hidden sector, and the dark energy field is the associated dark axion. The time-variation of the dark matter mass results in an effective dark energy equation of state that exhibits a phantom crossing behavior consistent with recent results. It also results in a slight delay in matter-radiation equality, which suppresses the overall growth of density perturbations.
\end{abstract}

\maketitle

\section{Introduction}

The fundamental natures of dark matter (DM) and dark energy (DE) remain two of the biggest mysteries in modern cosmology. To complicate matters, recent developments have added to the puzzles posed by these dominant components of the universe. 

Baryon Acoustic Oscillation (BAO) measurements from the Dark Energy Spectroscopic Instrument (DESI)~\cite{DESI:2024mwx}, together with observations of type Ia Supernovae (SN) from the Union3~\cite{Rubin:2023ovl} and Dark Energy Survey Year-5 data~\cite{DES:2024jxu}, have provided hints that DE is evolving over the history of the universe. In particular, the data prefer a DE equation of state $w < -1$ at high redshift and~$>-1$ today, indicating a ``phantom crossing" behavior~\cite{Feng:2004ad,Hu:2004kh,Guo:2004fq}. 
The very recent DESI DR2 results~\cite{desicollaboration2025desidr2resultsii,desicollaboration2025extendeddarkenergyanalysis} provide clearer and more robust evidence for this behavior, though see~\cite{Anselmi:2018vjz,ODwyer:2019rvi,Anselmi:2022exn} for possible caveats.

Meanwhile, the clustering of matter, as measured through weak lensing in galaxy surveys in the late universe~\cite{DES:2021wwk,KiDS:2020suj,Heymans:2020gsg,Dalal:2023olq,Amon:2022ycy}, seems to be less effective than that inferred from measurements of the Cosmic Microwave Background (CMB) in the early universe~\cite{Planck:2018vyg}. This discrepancy, usually called~$S_8$ tension, has persisted over several years at a statistical significance of about~$2-3\sigma$. Intringuingly, a simple modification to the late-time growth can resolve this tension~\cite{Lin:2023uux} (see also~\cite{Poulin:2022sgp,Nguyen:2023fip,Sabogal:2024yha}), with the modification taking place around the same time as the DE phantom crossing suggested by DESI.

Theoretically well-behaved and natural models of DE with~$w < -1$~\cite{Caldwell:1999ew} are challenging to construct. Indeed, the required violation of the Null Energy Condition almost invariably implies ghost/gradient instabilities~\cite{Carroll:2003st,Cline:2003gs,Dubovsky:2005xd} and/or superluminal propagation~\cite{Nicolis:2009qm,Creminelli:2010ba,Creminelli:2012my}. 

One should stress, however, that the evidence for~$w < -1$ crucially hinges on DESI's adopted parameterization for~$w(a)$~\cite{Chevallier:2000qy,Linder:2002et}, as well as the assumption of a non-interacting DM model.
Alternatively, the inferred phantom behavior may be due to non-minimal coupling of DE to gravity~\cite{Carroll:2004hc,Martin:2005bp,Ye:2024ywg,Yao:2025wlx}, or interactions between DM and DE~\cite{Huey:2004qv,Das:2005yj,Agrawal:2019dlm,Chakraborty:2025syu,Linder:2025zxb,Bedroya:2025fwh}. The latter possibility, which has a long history ({\it e.g.},~\cite{Amendola:1999er,Copeland:2006wr,Bolotin:2013jpa,Wang:2016lxa,Gomez-Valent:2020mqn,Rahimy:2025iyj,Giare:2024smz,Sabogal:2025mkp}), is the focus of this Letter.

Interacting DMDE theories generally suffer from a problem of technical naturalness. The DE scalar must be ultra-light to evolve dynamically in the late universe, but its coupling to DM implies large radiative corrections to its mass. Axions are a notable exception, thanks to their discrete shift symmetry. However, it is well-known that to successfully drive cosmic acceleration~\cite{Freese:1990rb,Frieman:1995pm} axions must either be placed near the maximum of their potential, which amounts to a tuning of initial conditions; or the axion decay constant must be super-Planckian, which runs afoul of the weak gravity conjecture~\cite{Banks:2003sx,Arkani-Hamed:2006emk}. 

In this Letter we present a well-motivated particle physics implementation of interacting DMDE with effective phantom crossing behavior, based on the mechanism of~\cite{Das:2005yj}, which is both technically natural and avoids tuning of initial conditions. The model assumes that DM is part of a ``dark QCD" sector~\cite{Dondi:2019olm,Morrison:2020yeg,Tsai:2020vpi,Garani:2021zrr}, with dark baryons serving as DM. Associated to this dark QCD sector is a dark axion~$\phi$, which serves as DE. 

In the presence of a finite density of dark baryons, the quark condensate is altered~\cite{Cohen:1991nk,Balkin:2020dsr}. This results in a density-dependent correction~\cite{Hook:2017psm} to the axion potential~\cite{DiVecchia:1980yfw,GrillidiCortona:2015jxo}, which encodes the DMDE interaction. Crucially, this allows the axion to successfully drive cosmic acceleration with a sub-Planckian decay constant, without tuning initial conditions.


Our framework shares some features with the axio-dilaton model of~\cite{Brax:2023qyp,Smith:2024ibv,Smith:2025grk}, but also important differences. In their case, the DE field (the dilaton) couples not only to DM (which can be the axion) but also to ordinary matter, potentially yielding distinct observational signatures, {\it e.g.}, in the CMB. As in our model, their axion potential acquires density-dependent corrections, here sourced by both dark and ordinary matter.

\vspace{-0.3cm}
\section{DMDE interactions}
\label{mech rev}

We begin by briefly summarizing the effects of a DMDE interaction, following~\cite{McDonough:2021pdg,Lin:2022phm}.
The full equations of motion for background and linear perturbations are given in Appendix~\ref{app: EoM}. 

The coupling between DM and a DE scalar field~$\phi$ results in DM having a~$\phi$-dependent mass
\begin{equation}
m(\phi) \equiv m_0 A(\phi)\,.
\label{Adef}
\end{equation}
Cosmologically, the DM number density still redshifts as~$1/a^3$, thanks to particle number conservation, but because of~\eqref{Adef} the DM mass/energy density is given by
\begin{equation}
    \rho_{\rm DM}(\phi) = \frac{A(\phi)}{A(\phi_0)}\frac{\rho_{\rm DM}^0}{a^3}\,,
    \label{eq:rho_DM}
\end{equation}
where~``0" indicates present-day values. 

The effective DE equation of state can then be derived as in~\cite{Das:2005yj}.
If one fits the data assuming a standard, non-interacting DM component and a decoupled DE component, then one is effectively ascribing the non-standard time evolution of DM to the DE density:
\begin{equation}
    \rho_{\rm DE}^{\rm eff} = \rho_{\phi} +\left [\frac{A(\phi)}{A(\phi_0)}-1\right] \frac{\rho_{\rm DM}^0}{a^3} \,,
\end{equation}
where~$\rho_\phi = \frac{1}{2a^2}\phi '^2 + V(\phi)$ is the scalar field density.
The corresponding effective DE equation of state is
\begin{equation}
    w_{\rm eff} = \frac{w_\phi}{1+\left[\frac{A(\phi)}{A(\phi_0)}-1\right]\frac{\rho_{\rm DM}^0}{a^3\rho_\phi}} \ ,
\label{eq:weff}
\end{equation}
where~$w_\phi = \frac{ \phi'^2 - 2a^2V(\phi)}{ \phi'^2 + 2a^2V(\phi)}$ is the standard scalar equation of state parameter. 
Notice that~$w_{\rm eff}$ and~$w_\phi$ coincide at present, but differ in the past. The DESI results suggest that~$w_{\rm eff} < -1$ in the past, but grew to a value~$w_{\rm eff} > -1$ at present. This can be achieved with~\eqref{eq:weff} provided that  
\begin{equation}\label{eq:A-requirement}
    \frac{A(\phi)}{A(\phi_0)} < 1
\end{equation}
at early times. This corresponds physically to a DM mass that increases with time in the DESI-sensitive epoch.\footnote{It is possible to achieve~$w_{\rm eff} < -1$ with a {\it decreasing} DM mass~\cite{Agrawal:2019dlm,Bedroya:2025fwh}, if~$\rho_{\rm DE}^{\rm eff}$ and~$\rho_\phi$ are chosen to coincide at an earlier redshift.}

The DMDE coupling also affects the growth of linear density inhomogeneities~$\delta_{\rm c}$ in three ways:
\vspace{-0.1cm}
\begin{itemize}
\item Since the DM mass increases with time, matter-radiation equality is shifted to a slightly later time. Because~$\delta_{\rm c}$ grows logarithmically in the radiation-dominated epoch and linearly in~$a$ in the matter-dominated epoch, the delay in matter-radiation equality suppresses the overall growth relative to~$\Lambda$CDM:
\begin{equation}\label{eq:deltac-ratio}
    \left.\frac{\delta_{\rm c}}{\delta_{\rm c}^{\Lambda{\rm CDM}}}\right\vert_{a=1}\simeq \frac{a_{\rm eq}^{\Lambda{\rm CDM}}}{a_{\rm eq}}\simeq  \frac{A(\phi_{\rm eq})}{A(\phi_0)}\, .
\end{equation}

\vspace{-0.1cm}
\item The scalar field changes the expansion history in the DE-dominated epoch. Compared to~$\Lambda$CDM, the background expansion with increasing~$w_{\rm eff}$ generically suppresses the matter density growth.

\vspace{-0.1cm}
\item The DE scalar field mediates an attractive fifth-force between DM, which enhances the density growth~\cite{McDonough:2021pdg}. However, because this effect is quadratic in the coupling~${\rm d}\ln A(\phi)/{\rm d}\phi$, it will be subdominant in our model compared to the previous two effects, which are linear in the coupling.

\end{itemize}

The general framework described above offers a simple mechanism to achieve~$w_{\rm eff} < -1$, with interesting phenomenological implications.
We now turn to implementing this simple idea in a well-motivated particle physics model that is stable, technically natural, and under calculational control.

\vspace{-0.3cm}
\section{Dark QCD Axion Model}
\label{dark QCD axion}

We consider the oft-studied idea that DM is part of a ``dark QCD" sector~\cite{Dondi:2019olm,Morrison:2020yeg,Tsai:2020vpi,Garani:2021zrr},
focusing on the simplest possibility of a dark~SU(3) gauge theory with two light (up/down) dark quarks, with masses~$m_{\rm u}$ and~$m_{\rm d}$. We assume that this sector couples to the visible sector only through gravity. Throughout the rest of the Letter we will sometimes just refer to quarks, baryons and pions, leaving implicit that they are dark sector objects.

Below the confinement scale, the relevant excitations are dark pions, which may be short-lived, and dark baryons, which are
accidentally long-lived thanks to approximate dark baryon number conservation~\cite{Garani:2021zrr}. For simplicity, we will assume  
that DM is comprised entirely of dark baryons.

A dark axion~$\phi$, associated with the dark QCD sector, acts as DE. Like the standard QCD axion, it acquires a vacuum potential due to the pion mass terms~\cite{DiVecchia:1980yfw,GrillidiCortona:2015jxo}:\footnote{Importantly, Eq.~\eqref{QCD pot} differs from the familiar~$V(\phi) = -\Lambda^4  \cos\frac{\phi}{2f}$ obtained in the single-instanton approximation~\cite{GrillidiCortona:2015jxo}. For equal quark masses~($\xi = 1$), Eq.~\eqref{QCD pot} reduces to~$V (\phi) \simeq - \Lambda^4\left\vert \cos \frac{\phi}{2f}\right\vert$.}
\begin{equation}
V (\phi) \simeq -\Lambda^4 \sqrt{1- \xi \sin^2\left(\frac{\phi}{2f}\right)}\,,
\label{QCD pot}
\end{equation}
where~$\xi = \frac{4m_{\rm u}m_{\rm d}}{(m_{\rm u}+m_{\rm d})^2}$ is the dimensionless ratio of quark masses.
We will take the axion decay constant~$f$ to lie in the range~$10^{-2} M_{\rm Pl} \lesssim f \lesssim M_{\rm Pl}$ preferred by UV considerations~\cite{Choi:1985bz,Banks:2003sx,Svrcek:2006yi,Hui:2016ltb}. 
 
The scale~$\Lambda$ is given in terms of the pion mass~$m_\pi$ and decay constant~$f_\pi$, and set to the DE~(meV) scale:\footnote{We cannot explain why~$\Lambda$ should be meV. That said, as in standard QCD,~$\Lambda$ is naturally much smaller than~$\Lambda_{\rm dQCD}$, which in turn is exponentially smaller than the Planck scale.}
\begin{equation}
\Lambda^4 = \epsilon m_\pi^2 f_\pi^2 \sim {\rm meV}^4\,.
\label{Lambda def}
\end{equation}
Following~\cite{Hook:2017psm}, we have introduced a parameter~$\epsilon \ll 1$, which, as we will see, allows finite-density corrections to compete with the vacuum potential~\eqref{QCD pot} within the perturbative regime.

The pion mass is related to the light quark masses via the Gell-Mann-Oakes-Renner (GOR) relation: 
\begin{equation}
\langle \bar{q}q \rangle_{\rm vac} = - \frac{m_\pi^2 f_\pi^2}{m_{\rm u} + m_{\rm d}} \,,
\label{GOR}
\end{equation}
where ``vac" indicates a vacuum expectation value. We henceforth take~$\langle \bar{u}u \rangle_{\rm vac} \simeq \langle \bar{d}d \rangle_{\rm vac}$, which is accurate up to small isospin symmetry breaking corrections. Expanding the potential~\eqref{QCD pot} to quadratic order around~$\phi = 0$, we obtain the axion mass:
\begin{equation}
m_\phi^2 =  \frac{\xi}{4} \frac{\Lambda^4}{f^2} = - \frac{\epsilon}{f^2} \frac{m_{\rm u} m_{\rm d}}{m_{\rm u} + m_{\rm d}} \langle \bar{u}u \rangle_{\rm vac}\,,
\end{equation}
where we have used Eqs.~\eqref{Lambda def}-\eqref{GOR}. With~$\epsilon = 1$ this is the well-known formula for the QCD axion mass~\cite{Weinberg:1977ma}.

When coupling to gravity, we should allow in general for a constant shift to the potential:
\begin{equation}
V(\phi) = \Lambda^4\left[1 + v - \sqrt{1- \xi \sin^2\left(\frac{\phi}{2f}\right)}\,\right]\,.
\label{vacuum pot}
\end{equation}
The dimensionless constant~$v$ sets the potential energy at the origin,~$V(0) = \Lambda^4 v$, and therefore is the cosmological constant. While we introduce this parameter for generality, our model does not rely on non-zero~$v$.

\vspace{-0.5cm}
\subsection{Effective axion coupling to DM}
\vspace{-0.2cm}

\begin{figure}
    \centering
    \includegraphics[width=0.99\linewidth]{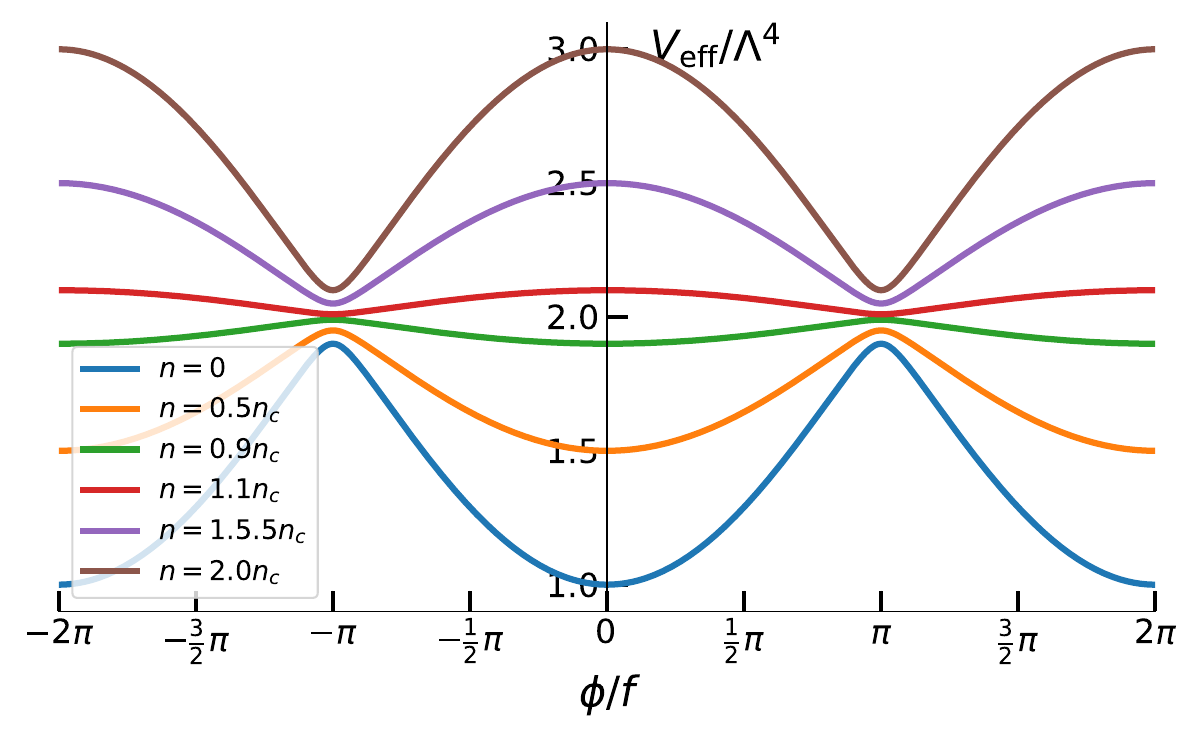}
    \caption{The dark axion effective potential (Eq.~\ref{axion pot full}) with different finite-density of dark baryons. Here we adopt $m_{\rm u}/m_{\rm d}=0.8$ and $v=1$ for illustration purposes. 
    }
    \label{fig:Veff}
\end{figure}

In the presence of a finite density of dark baryons, the quark condensate is altered~\cite{Cohen:1991nk,Balkin:2020dsr}:
\begin{equation}
\frac{\langle \bar{q}q \rangle_n}{\langle \bar{q}q \rangle_{\rm vac}} = 1 + \frac{1}{\langle \bar{q}q \rangle_{\rm vac}} \frac{\partial \Delta E}{\partial m_q}\,,
\label{cond density dep}
\end{equation}
where~$\Delta E$ is the energy shift of the QCD ground state at finite density, and the subscript ``$n$" denotes the finite number density of dark baryons. Ignoring nucleon interactions, such that dark nucleons behave as a free Fermi gas, and in the non-relativistic approximation, the energy shift is just given by the mass density $\Delta E \simeq  m n$.
%
%
It follows that
\begin{equation}
\frac{\langle \bar{q}q \rangle_n}{\langle \bar{q}q \rangle_{\rm vac}} \simeq 1 +  \frac{n\sigma_{\rm N}}{m_q \langle \bar{q}q \rangle_{\rm vac}}  \simeq 1 - \frac{2n\sigma_{\rm N}}{m_\pi^2 f_\pi^2} \,.
\label{cond at finite n}
\end{equation}
Here, the pion-nucleon sigma term~$\sigma_{\rm N} =  \frac{\partial m}{\partial\ln \bar{m}_{q}}$, with~$\bar{m}_q = \frac{1}{2} (m_{\rm u} + m_{\rm d})$, characterizes the dependence of the dark nucleon mass on the average quark mass.\footnote{\label{fotenote:param}For simplicity, we neglect the contribution from~$\tilde{\sigma}_N =\frac{\partial m}{\partial\ln \Delta m_q}$, which encodes the nucleon mass dependence on the quark mass difference~$\Delta m_q =\frac{1}{2}(m_{\rm u} - m_{\rm d})~$~\cite{Balkin:2020dsr}. For consistency, one should therefore focus on the limit of nearly equal quark masses, as we will do in the examples below.} It has units of mass, and for reference takes the value~$\sigma_{\rm N} \simeq 59\pm 7$~MeV~\cite{Alarcon:2011zs} in the Standard Model. It is important to stress that Eq.~\eqref{cond at finite n} is valid to linear order in the density, {\it i.e.}, neglecting~${\cal O}\left( \frac{n^2\sigma_{\rm N}^2}{m_\pi^4f_\pi^4} \right)$ corrections.

As first observed in~\cite{Hook:2017psm}, the density dependence of the quark condensate implies a finite-density correction~$U(\phi)$ to the axion potential. Combining Eqs.~\eqref{GOR} and~\eqref{cond at finite n}, we obtain
\begin{equation}
U(\phi) \simeq 2 \sigma_{\rm N}  n  \sqrt{1- \xi \sin^2\left(\frac{\phi}{2f}\right)} + {\cal O}\left( \frac{n^2\sigma_{\rm N}^2}{m_\pi^4f_\pi^4} \right) \,.
\label{U(a)}
\end{equation}
A few important points:

\vspace{-0.1cm}
\begin{itemize}

\item Since~$\sigma_{\rm N}$ is positive (heavier quarks means heavier nucleons),~$U(\phi)$ has the opposite sign to the square-root term in~$V(\phi)$. Hence minima of~$V(\phi)$ are maxima of~$U(\phi)$, and vice versa.

\vspace{-0.1cm}
\item From the point of view of DM, Eq.~\eqref{U(a)} implies an axion-dependent DM mass of the form~\eqref{Adef}, with 
\begin{equation}
A(\phi) \simeq 1 + 2\frac{\sigma_{\rm N}}{m_0} \sqrt{1- \xi\sin^2\left(\frac{\phi}{2f}\right)}\,.
\label{DM mass axion}
\end{equation}
Consequently, the dark axion mediates a force between dark baryons. Notice, however, that at the linearized level, the coupling between axion perturbations and DM is proportional to~$A_{,\phi}$, and therefore vanishes at minima/maxima of~$U$. 

\end{itemize}

Combining Eqs.~\eqref{vacuum pot} and~\eqref{U(a)}, we obtain the finite-density potential
\begin{align}
\nonumber
V_{\rm eff} (\phi) & =   \Lambda^4 \Bigg\{1 + v  - \left(1 - \frac{2\sigma_{\rm N} n}{\Lambda^4}\right)  \sqrt{1-\xi \sin^2\left(\frac{\phi}{2f}\right)}  \Bigg\} \\
& + {\cal O}\left( \frac{n^2\sigma_{\rm N}^2}{m_\pi^4f_\pi^4} \right)\,.
\label{axion pot full}
\end{align}
Notice that the coefficient of the square root term in~$V_{\rm eff}(\phi)$ can flip sign, depending on the DM density. This happens at a critical DM density~$n_{\rm c}$ given by
\begin{equation}
n_{\rm c} = \frac{\Lambda^4}{2\sigma_{\rm N}} = \frac{\epsilon m_\pi^2 f_\pi^2}{2\sigma_{\rm N}} \,.
\label{nc}
\end{equation}
Since~$\epsilon \ll 1$, the sign flip can occur in the regime~$\rho < \frac{m_\pi^2f_\pi^2}{2\sigma_N/m_0}$ where corrections in Eq.~\eqref{axion pot full} can be neglected~\cite{Hook:2017psm}. For a given~$\rho$, using Eq.~\eqref{Lambda def} this requires~$\epsilon < \frac{{\rm meV}^4}{2\rho \sigma_N/m_0}$. Demanding perturbative control up to~$z = 10^4$, say, requires $\epsilon \lesssim 10^{-12}$, where we have assumed~$\sigma_N/m_0 \lesssim 1$ for simplicity. Combined with Eq.~\eqref{Lambda def}, this implies
\begin{equation}
m_\pi f_\pi \gtrsim {\rm eV}^2\,.
\label{mpifpi bound}
\end{equation}
The required small~$\epsilon$ values can be achieved naturally with multiple DM copies with~$\mathds{Z}_N$ exchange symmetry~\cite{Hook:2018jle,DiLuzio:2021pxd}. We leave a detailed investigation of this scenario to future work.

Above critical density~($n > n_{\rm c}$), the potential is minimized at~$\phi = \pi f$; instead, at low density~($n < n_{\rm c}$), it is minimized at~$\phi = 0$. See Fig.~\ref{fig:Veff}. From Eq.~\eqref{DM mass axion}, this implies that the DM mass increases in time, as desired to achieve~$w_{\rm eff} < -1$. 

For the standard QCD axion, the finite-density correction can compete with the vacuum potential only at nuclear densities, {\it e.g.}, in the core of neutron stars~\cite{Hook:2017psm,Zhang:2021mks}. In our case, the scale of interest is~$\Lambda \sim {\rm meV}$ for the dark axion to act as DE,
hence the phase transition occurs at cosmological densities in the late universe.

\vspace{-0.5cm}
\subsection{Some remarks}
\vspace{-0.2cm}

We close this section with a few comments.

\vspace{0.1cm}
A first remark pertains to the chiral symmetry breaking phase transition, when dark quarks confine to form dark pions and baryons. 
As in QCD, this occurs at the scale~$\Lambda^3_{\rm dQCD}\sim \langle \bar{q}q \rangle_{\rm vac}$. Since~$m_{u,d} \ll \Lambda_{\rm dQCD}$, the GOR relation~\eqref{GOR} combined with~\eqref{mpifpi bound} implies~$\Lambda_{\rm dQCD} \gg {\rm eV}$, such that the phase transition occurs well before matter-radiation equality.\footnote{If baryons are fermions (with 3 colors), then the Tremaine-Gunn bound~\cite{Tremaine:1979we} requires~$m_0 \sim  \Lambda_{\rm dQCD} \gtrsim 10^2\,{\rm eV}$, which is easily satisfied for sufficiently light quarks. This bound is evaded entirely if baryons are scalars (with 2 colors).} Although this could have interesting cosmological implications, we will ignore the phase transition in studying the late-time dynamics. 
Since~$f_\pi \sim \Lambda_{\rm dQCD}\gg {\rm eV}$, it follows from~\eqref{mpifpi bound} that our dark pions are ultra-light, and thus dark baryons must comprise all of the DM. For dark baryons, the main constraint is the cosmological stability bound,~$m_0\sim \Lambda_{\rm dQCD} \lesssim 10^8\,{\rm GeV}$~\cite{Garani:2021zrr}. Ultra-light pions must satisfy the CMB bound on~$\Delta N_{\rm eff}$ (see Eq.~(6) of~\cite{Garani:2021zrr}), which is easily satisfied with small or even moderate ratio~$\xi$ of DM to SM temperatures at the phase transition~\cite{Garani:2021zrr}.


It is well known that, to act as DE, either the axion must be placed initially near the maximum of its potential~($\phi = \pi f$), which amounts to a tuning of initial conditions; or~$f$ must take on super-Planckian values~\cite{Freese:1990rb,Frieman:1995pm}, which is believed to be impossible in consistent theories of quantum gravity~\cite{Banks:2003sx,Arkani-Hamed:2006emk}. In our case, the axion naturally finds itself near~$\pi f$ by the onset of cosmic acceleration, since the density-dependent term~$U(\phi)$, which dominates until then, is minimized there.

\vspace{-0.5cm}
\section{Working Examples}
\label{working egs}

\begin{figure}
    \centering
    \includegraphics[width=0.99\linewidth]{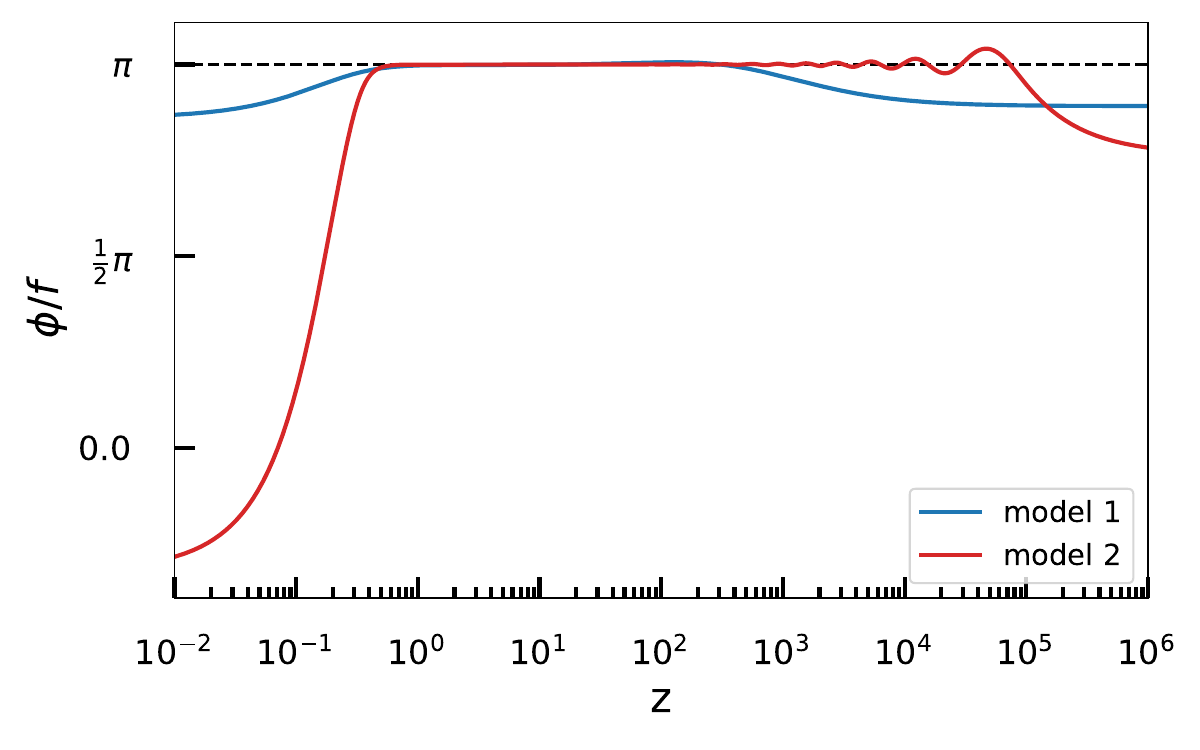}
    \caption{Scalar field evolution as a function of redshift for the two fiducial models. 
    }
    \label{fig:phi}
\end{figure}

\begin{figure}
    \centering
    \includegraphics[width=0.99\linewidth]{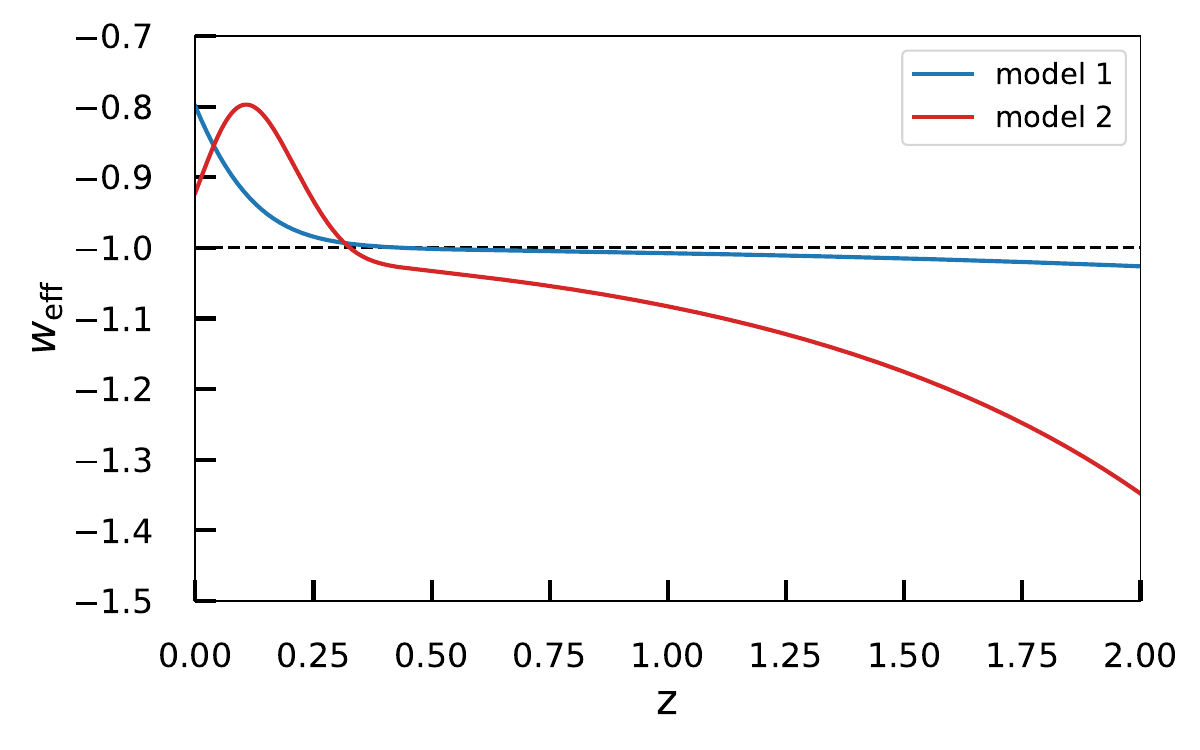}
    \caption{Effective DE equation of state,~$w_{\rm eff}$, defined in Eq.~\ref{eq:weff}, as a function of redshift for two working examples.
    }
    \label{fig:weff}
\end{figure}

\begin{figure}
    \centering
    \includegraphics[width=0.99\linewidth]{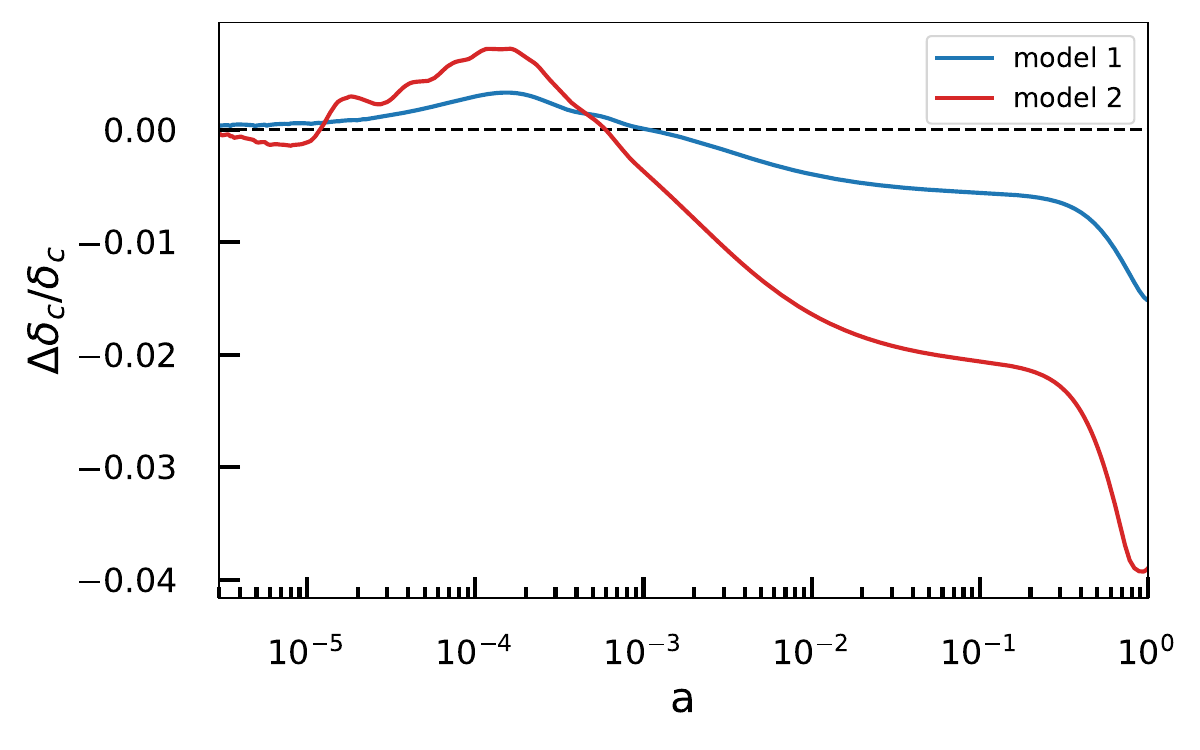}
    \caption{Fractional change to the DM density contrast relative to~$\Lambda$CDM,~$\frac{\Delta \delta_{\rm c}}{\delta_{\rm c}} = \frac{\delta_{\rm c} - \delta_{\rm c}^{\Lambda {\rm CDM}}}{\delta_{\rm c}^{\Lambda {\rm CDM}}}$, as a function of the scale factor.}
    \label{fig:deltac}
\end{figure}

In this section we provide two working examples with typical parameter values (for details, see Appendix~\ref{app: param variations}), leaving full likelihood analyses for future work. We compute background cosmological solutions and the evolution of linear perturbations using a modified version of CLASS~\cite{2011arXiv1104.2932L,2011JCAP...07..034B}.
We will see that our model can exhibit a phantom-crossing expansion history, consistent with that suggested by DESI, while simultaneously alleviating the~$S_8$ tension.

\vspace{-0.5cm}
\subsection{Background Cosmology}
\vspace{-0.2cm}

Figure~\ref{fig:phi} shows the behavior of the axion field as a function of redshift for the two fiducial models. At early times, the density-dependent term~$U(\phi)$ dominates, and the field rapidly settles to the minimum of the effective potential at~$\phi = \pi f$. Thus, for a broad range of initial conditions, the axion naturally achieves the necessary conditions to drive cosmic acceleration at later times. This feature is critically important, since it allows us to restrict ourselves to sub-Planckian values of~$f$ while avoiding tuning of initial conditions. 

As the DM density drops below the critical density~\eqref{nc}, the vacuum term~$V(\phi)$ begins to dominate, and the minimum at~$\phi=\pi f$ becomes a maximum. This potential flip happens at~$z\simeq7$ and~$z\simeq2.5$ in Models~1 and~2, respectively.
The field then begins to roll towards either the~$\phi>\pi f$ or~$<\pi f$ direction, depending on small variations of the field value at the critical density. Whichever direction the field rolls towards, the resulting physics is equivalent due to the periodicity of the potential.

Figure~\ref{fig:weff} shows the effective DE equation of state, defined in Eq.~\eqref{eq:weff}, for the two fiducial models. As expected, we see that~$w_{\rm eff}<-1$ due to the evolving DM mass, while at present~$w_{\rm eff}\sim-0.8$. The phantom crossing occurs at~$z\simeq 0.5$ and~$\simeq 0.25$ in Models~1 and~2, respectively. When mapping to the widely-used parameterization~$w(a)=w_0+(1-a)w_a$~\cite{Chevallier:2000qy,Linder:2002et}, this leads to $w_0>-1$ and $w_a<0$.
See Appendix~\ref{app: param variations} for more details on the sensitivity of observables to variations in parameter values.

\vspace{-0.5cm}
\subsection{Evolution of Perturbations}\label{sec:perturbation}
\vspace{-0.2cm}

We now consider the behavior of linear perturbations. Figure~\ref{fig:deltac} shows the fractional change~$\frac{\Delta \delta_{\rm c}}{\delta_{\rm c}} = \frac{\delta_{\rm c} - \delta_{\rm c}^{\Lambda {\rm CDM}}}{\delta_{\rm c}^{\Lambda {\rm CDM}}}$ for the two fiducial models relative to the~$\Lambda$CDM model.

The evolution of~$\frac{\Delta \delta_{\rm c}}{\delta_{\rm c}}$ shows three distinct effects. Before matter-radiation equality, the density contrast is temporarily larger than in~$\Lambda$CDM, as the field settles to the minimum of the effective potential. From matter-radiation equality onwards,~$\delta_{\rm c}$ monotonically drops below the~$\Lambda$CDM value. The suppression during the matter-dominated epoch is due to the shift in matter-radiation equality (see Eq.~\ref{eq:deltac-ratio}), while the further suppression during the DE-dominated epoch is due to DE having a different equation of state than in~$\Lambda$CDM. Since the net effect is a suppression in~$\delta_{\rm c}$, this confirms that the attractive scalar-mediated force between DM particles is subdominant.

Quantitatively, we see from Fig.~\ref{fig:deltac} that the growth of matter overdensities result in a~$\simeq 1.5-4\%$ suppression in the DM density fluctuation today compared to~$\Lambda$CDM. The resulting~$S_8$ values are~$0.793$ and~$0.776$ in Models~1 and~2, respectively, compared to~$0.809$ in the~$\Lambda$CDM model.
Thus, the model simultaneously addresses the behavior of DE suggested by DESI and reduces the~$S_8$ tension.

\section{Conclusions and Discussions}
\label{conclu sec}

In this Letter we have presented a theoretically well-motivated model of interacting DM and DE, which naturally gives rise to an effective DE equation of state~$w_{\rm eff} < -1$ based on the mechanism of~\cite{Das:2005yj}. The model capitalizes on the well-studied idea that DM is part of a strongly-interacting dark QCD sector, while DE is played by a dark axion associated with this sector. The density dependence of the axion potential embodies a technically natural interaction between the DM and DE components. 

For the purpose of this preliminary study, we have provided two illustrative examples with natural values of parameters. In both cases, the effective DE equation of state exhibits a phantom crossing behavior, consistent with the recent evidence from DESI BAO and SNIa data sets. Furthermore, the DMDE coupling results in a slight suppression in the growth of structures, which can offer a resolution to the~$S_8$ tension. 

In future work, we will perform full likelihood analyses of this model. It will be interesting to explore other phenomenological implications of the scenario, in particular for the Hubble tension~\cite{Freedman:2017yms,Verde:2019ivm,Kamionkowski:2022pkx}, {\it e.g.}~\cite{Pitrou:2023swx}.
Although the axion-DM coupling could in principle give rise to ``screening"~\cite{Hook:2017psm,Brax:2023qyp} in high-density regions, akin to the symmetron mechanism~\cite{Hinterbichler:2010es}, as shown in Appendix~\ref{app: screening} cosmic structures remain unscreened in our parameter regime.

\textbf{Acknowledgements:} 
We thank Cliff Burgess, Colin Hill, Lam Hui, Keisuke Harigaya, and Eric Linder for helpful discussions, and especially Bhuvnesh Jain for collaboration in the initial stages of this work and many insightful discussions. The work of J.K. and M.T. is supported in part by the DOE (HEP) Award DE-SC0013528. M-X.L. is supported by funds provided by the Center for Particle Cosmology.

\bibliography{ref} 

\clearpage

\appendix

\begin{figure}
    \centering
    \includegraphics[width=0.99\linewidth]{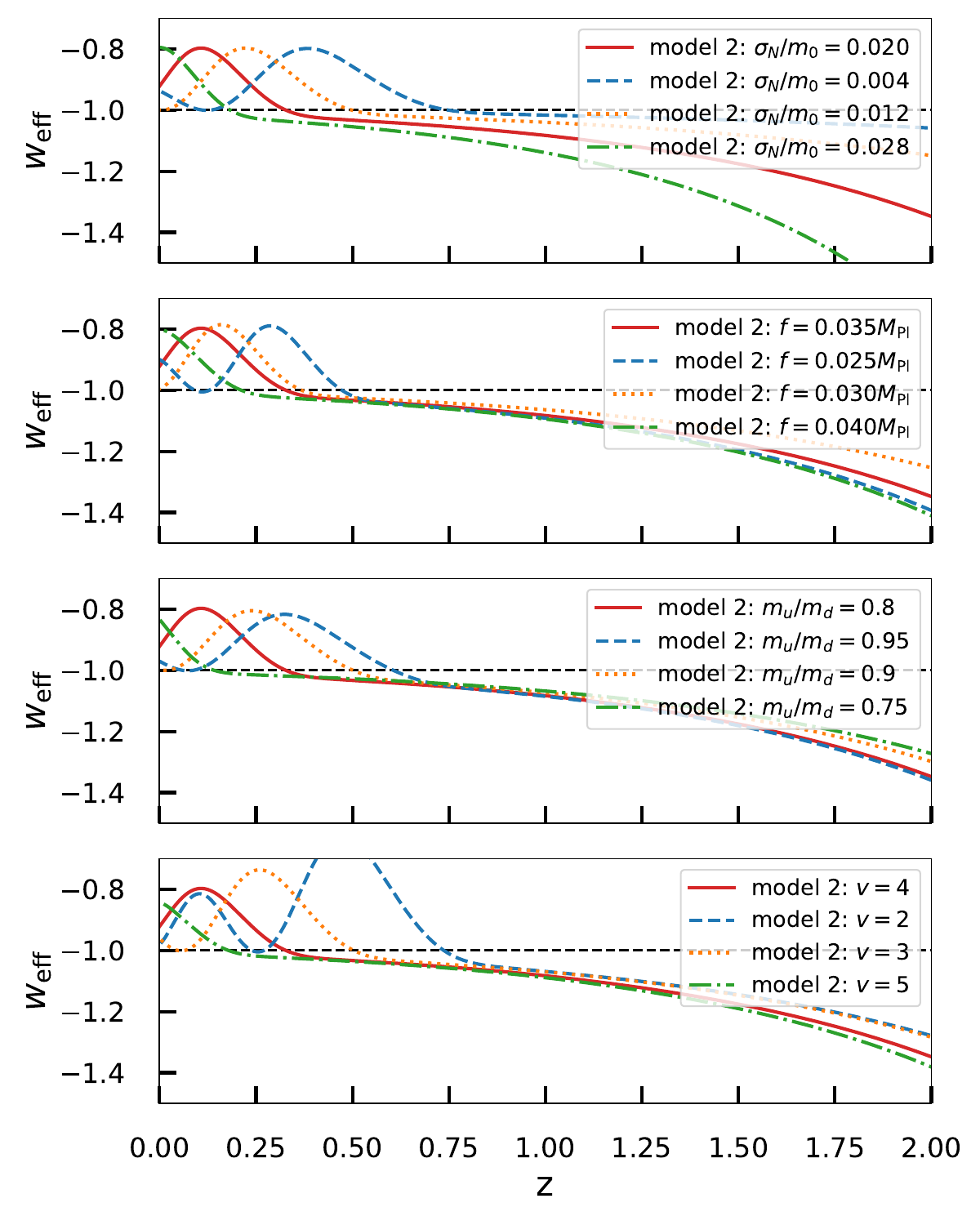}
    \caption{Impact on the effective DE equation of state~$w_{\rm eff}$ on varying the model parameters around our fiducial example Model 2 (red line).
    }
    \label{fig:weff-vary}
\end{figure}

\begin{figure}
    \centering
    \includegraphics[width=0.99\linewidth]{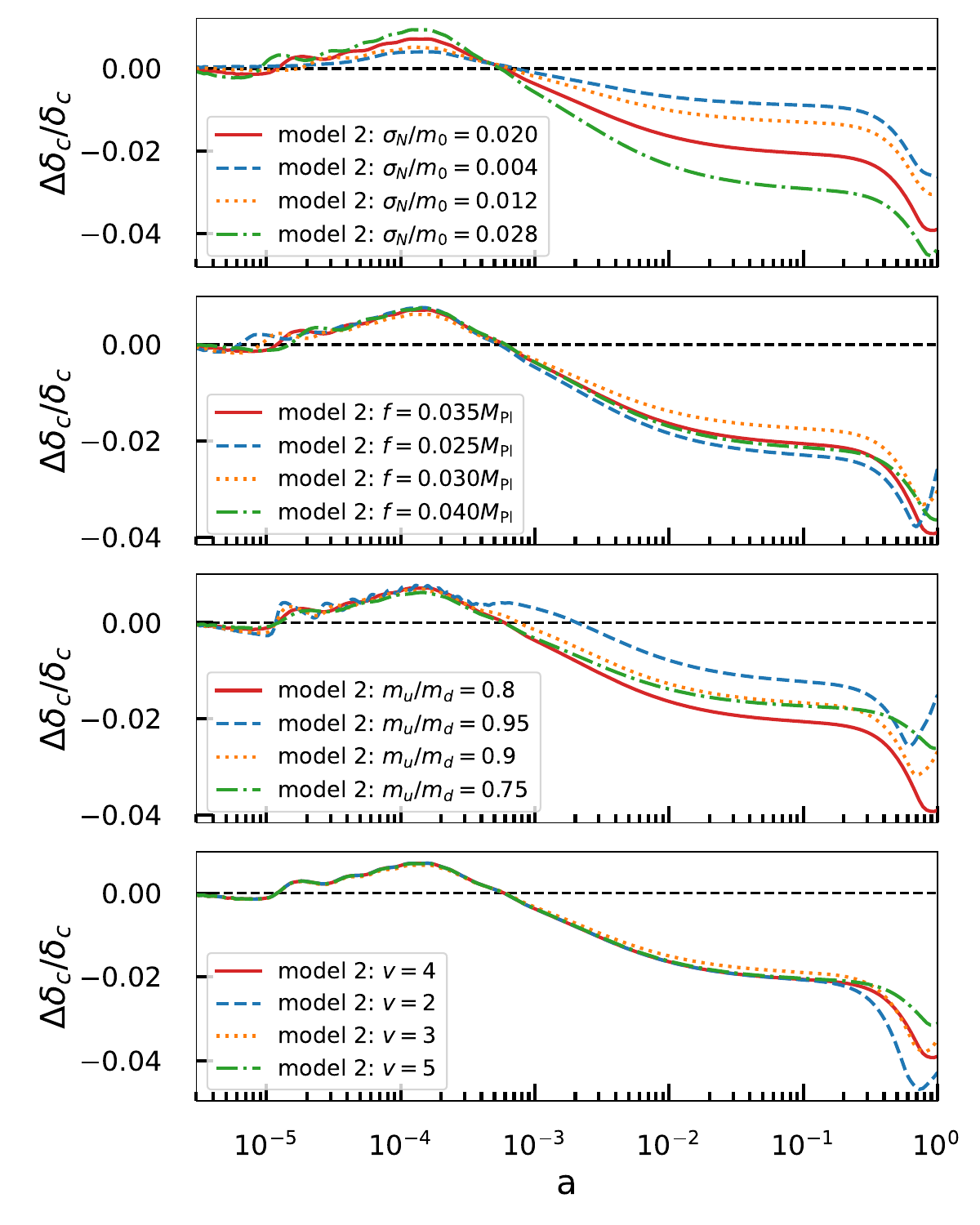}
    \caption{Impact on the DM density contrast~$\Delta\delta_{\rm c}/\delta_{\rm c}$ of varying the model parameters around our fiducial example Model 2 (red line).
    }
    \label{fig:deltac-vary}
\end{figure}

\section{Equations of Motion for DMDE interactions}
\label{app: EoM}

Here we provide the equations of motion for the background and linear perturbations following~\cite{McDonough:2021pdg,Lin:2022phm}. Consider a Lagrangian 
\begin{equation}
    \mathcal{L} = -\frac{1}{2} (\partial \phi)^2+ {\rm i}\bar{\psi} \slashed{D}\psi  - V(\phi) - m_0 A(\phi)\bar{\psi}\psi \,,
    \label{eq:Lmodel}
\end{equation}
where~$\phi$ is the DE scalar field with potential~$V(\phi)$, and~${\psi}$ is a fermionic DM candidate with~$\phi$-dependent mass given by Eq.~\eqref{Adef}. On a cosmological background with cosmic scale factor~$a$, the equation of motion for~$\phi$ in the presence of an ambient matter density~$\rho_{\rm DM}$ is
\begin{equation}
    \phi '' + 2aH\phi ' = -a^2 \left(\frac{{\rm d}V}{{\rm d}\phi} + \frac{{\rm d}\ln A(\phi)}{{\rm d}\phi}\rho_{\rm DM}(\phi)\right) \,,
\end{equation}
where primes denote derivatives with respect to conformal time~$\tau$, and~$H= a'/a^2$ is the Hubble parameter. 
As usual, conformal time is related to cosmic proper time~$t$ by~${\rm d}\tau={\rm d}t/a(t)$.
The quantity in parentheses on the right-hand side is naturally interpreted as the derivative~${\rm d}V_{\rm eff}/{\rm d}\phi$ of an effective, density-dependent potential 
\begin{equation}
    V_{\rm eff}(\phi) = V(\phi) + \rho_{\rm DM}(\phi)\,.
\end{equation}
Because of its coupling to~$\phi$, the DM density conservation equation is modified to
\begin{equation}
    \rho_{\rm DM}' +3aH\rho_{\rm DM} =  \phi' \frac{{\rm d}\ln A(\phi)}{{\rm d}\phi} \rho_{\rm DM} \,.
\end{equation}
This can readily be integrated:
\begin{equation}
    \rho_{\rm DM}(\phi) = \frac{A(\phi)}{A(\phi_0)}\frac{\rho_{\rm DM}^0}{a^3}\,,
\end{equation}
where~``0" indicates present-day values, and we have set~$a_0 = 1$ without loss of generality. 
The corresponding Friedmann equation is
\begin{equation}
    3H^2M_{\rm Pl}^2 = \rho_\phi + \rho_{\rm DM}(\phi)\,,
\end{equation}
where $M_{\rm Pl} = \frac{1}{\sqrt{8\pi G_{\rm N}}}$ is the reduced Planck mass, and the scalar field energy density is 
\begin{equation}
    \rho_\phi = \frac{1}{2a^2}\phi '^2 + V(\phi) \,.
\end{equation}

The equations of motion for linear perturbations in synchronous gauge are given by
\begin{eqnarray}
\delta\phi'' +2aH\delta\phi' +\left(k^2+a^2\frac{{\rm d}^2V}{{\rm d}\phi^2}\right)\delta\phi +\frac{1}{2}h'\phi' = \nonumber\\
    -a^2 \left[\frac{{\rm d}\ln A(\phi)}{{\rm d}\phi}\delta_{\rm c} +\frac{{\rm d}^2\ln A(\phi)}{{\rm d}\phi^2}\delta\phi\right]\rho_{\rm DM}(\phi)\,;
\label{pert 1}
\end{eqnarray}
\begin{equation}
    \delta_{\rm c}'+\theta+\frac{h'}{2} = \frac{{\rm d}\ln A(\phi)}{{\rm d}\phi}\delta\phi' +\frac{{\rm d}^2\ln A(\phi)}{{\rm d}\phi^2}\phi'\delta\phi \,;
\label{pert 2}
\end{equation}
\begin{equation}
    \theta'+aH\theta = \frac{{\rm d}\ln A(\phi)}{{\rm d}\phi}k^2\delta\phi -\frac{{\rm d}\ln A(\phi)}{{\rm d}\phi}\phi'\theta \,,
\label{pert 3}
\end{equation}
where~$\delta_{\rm c}$ and~$\theta\equiv\partial_i v^i$ are the DM density and velocity perturbations, respectively, and~$h$ is the metric trace perturbation in synchronous gauge.

\section{Model parameters and their impacts on observables}
\label{app: param variations}

\vspace{0.1cm}
Our model is characterized by 5 free parameters:

\vspace{-0.2cm}
\begin{itemize}

\item The overall scale~$\Lambda$ of the axion potential, which we take to be~$\sim {\rm meV}$ to achieve late-time acceleration.

\vspace{-0.1cm}
\item The axion decay constant~$f$ and quark mass ratio~$m_{\rm u}/m_{\rm d}$ (equivalently,~$\xi$). Together, these parameters determine the curvature around the maximum/minimum of the potential, and therefore control the rate at which the axion rolls down. In practice we will take~$f \lesssim M_{\rm Pl}$ and~$m_{\rm u} \lesssim m_{\rm d}$.

\vspace{-0.1cm}
\item The parameter~$v$, which together with~$\Lambda$ sets the vacuum energy at~$\phi = 0$. We will take~$v \sim {\cal O}(1)$, compatible with~$V(\phi = 0) \sim {\rm meV}^4$. 

\vspace{-0.1cm}
\item The coupling parameter~$\sigma_{\rm N}/m_0$, which from Eq.~\eqref{nc} sets the critical DM density at which the axion becomes unstable. For context, in the Standard Model this takes the value~$\frac{\sigma_{\rm N}}{m_{\rm p}} = \frac{59~{\rm MeV}}{938~{\rm MeV}}\simeq 0.06$. 

\end{itemize} 

For illustrative purposes, we consider 2 fiducial models:

\vspace{-0.1cm}
\begin{itemize}

\item {\bf Model 1} has~$\frac{\sigma_{\rm N}}{m_0} = 0.01$,~$f = 0.25\, M_{\rm Pl}$, and~$v = 0$. 

\vspace{-0.1cm}
\item {\bf Model 2} has~$\frac{\sigma_{\rm N}}{m_0} = 0.02$,~$f = 0.035\, M_{\rm Pl}$, and~$v = 4$.  

\end{itemize}

\vspace{-0.1cm}
\noindent In both cases, we take~$m_{\rm u} = 0.8\, m_{\rm d}$, consistent with the discussion in Footnote~\ref{fotenote:param}. Interestingly, the assumed values for~$\sigma_{\rm N}/m_0$ are not dissimilar to the Standard Model value of~$0.06$, as mentioned above.

When comparing to~$\Lambda$CDM predictions, we must specify which observables to keep fixed. We choose to fix the angular size~$\theta_s$ of the sound horizon at the last scattering surface, as this is directly constrained by the CMB, as well as the fractional densities in DM and baryons today~$\Omega_{\rm c}$ and~$\Omega_{\rm b}$, as BAO and SN constrain~$\Omega_{\rm c}+\Omega_{\rm b}$. We set these parameters to Planck best-fit $\Lambda$CDM values~\cite{Planck:2018vyg}. Consequently,~$H_0$ is not fixed in the comparison. 

We also explore the impact on the expansion and growth histories of varying the parameters. Figures~\ref{fig:weff-vary} and \ref{fig:deltac-vary} show the changes in~$w_{\rm eff}$ and~$\Delta\delta_{\rm c}/\delta_{\rm c}$, respectively, around on Model 2. 

The coupling~$\sigma_{\rm N}/m_0$ controls the redshift at which~$w_{\rm eff}$ crosses~$-1$, with larger~$\sigma_{\rm N}/m_0$ corresponding to later crossing time. The potential energy at the origin, set by~$v$, controls the final value of~$w_{\rm eff}$, with~$w_{\rm eff}$ approaching~$-1$ for larger~$v$. As mentioned earlier, the axion decay constant~$f$ and quark mass ratio~$m_{\rm u}/m_{\rm d}$ control the curvature of the potential, and therefore how quickly the field rolls towards the minimum. Given the relatively limited constraining power of current observational data, we expect some partial degeneracy between the parameters. 

\section{Screening mechanism}\label{app: screening}

Because its potential depends on the DM density, the axion can in principle develop a spatially-dependent profile within sufficiently dense cosmic structures. Such structures thus source the axion, or equivalently are said to be ``screened" in the language of screening mechanisms.

For simplicity, consider a spherically symmetric halo of radius~$R$ and homogeneous density~$\rho$ well above the critical density given in Eq.~\eqref{nc}. As shown in~\cite{Hook:2017psm}, sourcing requires
\begin{equation}
R \gtrsim m_{\rm eff}^{-1}\,,
\label{screen cond}
\end{equation}
where~$m_{\rm eff}$ is the effective axion mass deep inside the screened halo:
\begin{equation}
m_{\rm eff}^2 = \left.\frac{{\rm d}^2 U(\phi)}{{\rm d}\phi^2}\right\vert_{\phi = \pi f} = \frac{\xi}{\sqrt{1-\xi}} \frac{\sigma_{\rm N}\rho}{2m_0f^2}\,.  
\label{meff explicit}
\end{equation}
In terms of the Newtonian potential~$\Phi = \rho R^2/6M_{\rm Pl}^2$ of the halo, Eqs.~\eqref{screen cond} and~\eqref{meff explicit} can be cast as
\begin{equation}
\Phi \gtrsim \frac{\sqrt{1-\xi}}{3\xi}\frac{m_0}{\sigma_{\rm N}}\frac{f^2}{M_{\rm Pl}^2}\,,
\label{screen cond 2}
\end{equation}
which is closely similar to the symmetron screening condition~\cite{Hinterbichler:2010es}. 

Virialized structures in the universe have characteristic potential~$\Phi\lesssim 10^{-6}-10^{-5}$. For the fiducial parameter values considered in this work, in particular~$10^{-2}\,M_{\rm Pl}\lesssim f \lesssim M_{\rm Pl}$, it is easy to see that Eq.~\eqref{screen cond 2} fails to be satisfied. That is, halos are not screened, {\it i.e.}, they do not source a non-linear inhomogeneous profile for the axion. That said, screening could be relevant for other regions of parameter space, {\it e.g.}, for applications to DM galactic phenomenology~\cite{Denisontoappear}.

\end{document}